\documentclass[12pt,a4paper]{article}

\usepackage{ifthen} 
\newboolean{pdflatex}
\setboolean{pdflatex}{true} 

\newboolean{articletitles}
\setboolean{articletitles}{true} 

\newboolean{uprightparticles}
\setboolean{uprightparticles}{false} 

\newboolean{inbibliography}
\setboolean{inbibliography}{false} 

\usepackage[top=1in, bottom=1.25in, left=1in, right=1in]{geometry}

\usepackage{microtype}
\usepackage{lineno}  
\usepackage{xspace} 
\usepackage{caption} 

\usepackage{graphicx}  
\usepackage{color}
\usepackage{colortbl}
\graphicspath{{./figs/}} 

\usepackage{amsmath} 
\usepackage{amssymb}
\usepackage{amsfonts}
\usepackage{upgreek} 

\newcommand*\patchAmsMathEnvironmentForLineno[1]{%
\expandafter\let\csname old#1\expandafter\endcsname\csname #1\endcsname
\expandafter\let\csname oldend#1\expandafter\endcsname\csname
end#1\endcsname
 \renewenvironment{#1}%
   {\linenomath\csname old#1\endcsname}%
   {\csname oldend#1\endcsname\endlinenomath}%
}
\newcommand*\patchBothAmsMathEnvironmentsForLineno[1]{%
  \patchAmsMathEnvironmentForLineno{#1}%
  \patchAmsMathEnvironmentForLineno{#1*}%
}
\AtBeginDocument{%
\patchBothAmsMathEnvironmentsForLineno{equation}%
\patchBothAmsMathEnvironmentsForLineno{align}%
\patchBothAmsMathEnvironmentsForLineno{flalign}%
\patchBothAmsMathEnvironmentsForLineno{alignat}%
\patchBothAmsMathEnvironmentsForLineno{gather}%
\patchBothAmsMathEnvironmentsForLineno{multline}%
\patchBothAmsMathEnvironmentsForLineno{eqnarray}%
}

\usepackage{hyperref}    
\hypersetup{
  colorlinks=true,
  linktoc=all,
  linkcolor=blue,
  citecolor=blue
}
\usepackage[all]{hypcap} 

\usepackage{xspace} 
\usepackage{upgreek}

\def\lhcb {\mbox{LHCb}\xspace}

\def\MagUp {\mbox{\em Mag\kern -0.05em Up}\xspace}

\ifthenelse{\boolean{uprightparticles}}%
{

 \def\PDelta      {\ensuremath{\Delta}\xspace}                 
 \def\PXi      {\ensuremath{\Xi}\xspace}                 
 \def\PLambda      {\ensuremath{\Lambda}\xspace}                 
 \def\PSigma      {\ensuremath{\Sigma}\xspace}                 
 \def\POmega      {\ensuremath{\Omega}\xspace}                 
 \def\PUpsilon      {\ensuremath{\Upsilon}\xspace}                 
 
 \def\PB      {\ensuremath{\mathrm{B}}\xspace}                 
                  
 \def\PD      {\ensuremath{\mathrm{D}}\xspace}

 \def\PK      {\ensuremath{\mathrm{K}}\xspace}

 \def\Pi      {\ensuremath{\mathrm{i}}\xspace}

}
{

 \mathchardef\PDelta="7101
 \mathchardef\PXi="7104
 \mathchardef\PLambda="7103
 \mathchardef\PSigma="7106
 \mathchardef\POmega="710A
 \mathchardef\PUpsilon="7107
                  
 \def\PB      {\ensuremath{B}\xspace}                 
                  
 \def\PD      {\ensuremath{D}\xspace}

 \def\PK      {\ensuremath{K}\xspace}

 \def\Pi      {\ensuremath{i}\xspace}

}

\makeatletter
\ifcase \@ptsize \relax
  \newcommand{\miniscule}{\@setfontsize\miniscule{4}{5}}
\or
  \newcommand{\miniscule}{\@setfontsize\miniscule{5}{6}}
\or
  \newcommand{\miniscule}{\@setfontsize\miniscule{5}{6}}
\fi
\makeatother

\DeclareRobustCommand{\optbar}[1]{\shortstack{{\miniscule (\rule[.5ex]{1.25em}{.18mm})}
  \\ [-.7ex] $#1$}}

  \def\Kbar    {{\kern 0.2em\overline{\kern -0.2em \PK}{}}\xspace}

\def\KorKbar    {\kern 0.18em\optbar{\kern -0.18em K}{}\xspace}

  \def\Dbar    {{\kern 0.2em\overline{\kern -0.2em \PD}{}}\xspace}

\def\DorDbar    {\kern 0.18em\optbar{\kern -0.18em D}{}\xspace}

\def\Bbar    {{\ensuremath{\kern 0.18em\overline{\kern -0.18em \PB}{}}}\xspace}

\def\BorBbar    {\kern 0.18em\optbar{\kern -0.18em B}{}\xspace}

  \def\Y#1S{\ensuremath{\PUpsilon{(#1S)}}\xspace}

\def\Lbar        {{\ensuremath{\kern 0.1em\overline{\kern -0.1em\PLambda}}}\xspace}
\def\LorLbar    {\kern 0.18em\optbar{\kern -0.18em \PLambda}{}\xspace}

\def\AT#1     {\ensuremath{A_{\mathrm{T}}^{#1}}\xspace}           

\def\C#1      {\ensuremath{\mathcal{C}_{#1}}\xspace}                       
\def\Cp#1     {\ensuremath{\mathcal{C}_{#1}^{'}}\xspace}                    
\def\Ceff#1   {\ensuremath{\mathcal{C}_{#1}^{\mathrm{(eff)}}}\xspace}        
\def\Cpeff#1  {\ensuremath{\mathcal{C}_{#1}^{'\mathrm{(eff)}}}\xspace}       
\def\Ope#1    {\ensuremath{\mathcal{O}_{#1}}\xspace}                       
\def\Opep#1   {\ensuremath{\mathcal{O}_{#1}^{'}}\xspace}                    

\newcommand{\tev}{\ifthenelse{\boolean{inbibliography}}{\ensuremath{~T\kern -0.05em eV}\xspace}{\ensuremath{\mathrm{\,Te\kern -0.1em V}}}\xspace}
\newcommand{\gev}{\ensuremath{\mathrm{\,Ge\kern -0.1em V}}\xspace}
\newcommand{\mev}{\ensuremath{\mathrm{\,Me\kern -0.1em V}}\xspace}
\newcommand{\kev}{\ensuremath{\mathrm{\,ke\kern -0.1em V}}\xspace}
\newcommand{\ev}{\ensuremath{\mathrm{\,e\kern -0.1em V}}\xspace}
\newcommand{\gevc}{\ensuremath{{\mathrm{\,Ge\kern -0.1em V\!/}c}}\xspace}
\newcommand{\mevc}{\ensuremath{{\mathrm{\,Me\kern -0.1em V\!/}c}}\xspace}
\newcommand{\gevcc}{\ensuremath{{\mathrm{\,Ge\kern -0.1em V\!/}c^2}}\xspace}
\newcommand{\gevgevcccc}{\ensuremath{{\mathrm{\,Ge\kern -0.1em V^2\!/}c^4}}\xspace}
\newcommand{\mevcc}{\ensuremath{{\mathrm{\,Me\kern -0.1em V\!/}c^2}}\xspace}

\def\invfb   {\ensuremath{\mbox{\,fb}^{-1}}\xspace}

\def\gsim{{~\raise.15em\hbox{$>$}\kern-.85em
          \lower.35em\hbox{$\sim$}~}\xspace}
\def\lsim{{~\raise.15em\hbox{$<$}\kern-.85em
          \lower.35em\hbox{$\sim$}~}\xspace}

\def\tell1  {TELL1\xspace}
\def\ukl1   {UKL1\xspace}

\usepackage{physics}

\usepackage{cite} 
\usepackage{mciteplus}

\usepackage{longtable} 

\usepackage{graphicx}
\usepackage{amsmath}

\usepackage{siunitx}
\DeclareSIUnit\evolt{eV}
\DeclareSIUnit[allow-number-unit-breaks]\neq{1~\mega\evolt~\text{neutron equivalent}\per\centi\meter\squared}

\usepackage[capitalize]{cleveref}
\usepackage{subcaption}
\usepackage{booktabs}
\usepackage{lineno}

\graphicspath{{./fig/}}

\newcommand\twgraphic[1]{\includegraphics[width=\textwidth]{#1}}

\title{Signal coupling to embedded pitch adapters in silicon sensors}

\usepackage{authblk}

\author[a]{M.~Artuso}
\author[b]{C.~Betancourt}
\author[b]{I.~Bezshyiko}
\author[a]{S.~Blusk}
\author[b]{R.~Brundler Denzer}
\author[c]{S.~Bugiel}
\author[c]{R.~Dasgupta}
\author[c]{A.~Dendek}
\author[d,e]{B.~Dey}
\author[a]{S.~Ely}
\author[b]{F.~Lionetto}
\author[d,e]{M.~Petruzzo}
\author[a]{I.~Polyakov}
\author[a]{M.~Rudolph}
\author[f]{H.~Schindler}
\author[b]{O.~Steinkamp}
\author[a]{S.~Stone}

\affil[a]{Syracuse University, Syracuse, NY USA}
\affil[b]{Universit\"{a}t Z\"{u}rich, Z\"{u}rich, Switzerland}
\affil[c]{AGH University of Science and Technology, Krak\'{o}w, Poland}
\affil[d]{Universit\`{a} degli Studi di Milano, Milan, Italy}
\affil[e]{Istituto Nazionale di Fisica Nucleare (INFN) - Sezione di Milano, Italy}
\affil[f]{European Organization for Nuclear Research (CERN), Geneva, Switzerland}

\makeatletter
\let\thetitle\@title
\let\theauthors\@author
\makeatother

\begin{document}

\renewcommand{\thefootnote}{\fnsymbol{footnote}}
\setcounter{footnote}{1}

\begin{titlepage}
\pagenumbering{roman}

\vspace*{-1.5cm}
\centerline{\large EUROPEAN ORGANIZATION FOR NUCLEAR RESEARCH (CERN)}
\vspace*{1.5cm}
\noindent
\begin{tabular*}{\linewidth}{lc@{\extracolsep{\fill}}r@{\extracolsep{0pt}}}
\ifthenelse{\boolean{pdflatex}}
{\vspace*{-2.7cm}\mbox{\!\!\!\includegraphics[width=.14\textwidth]{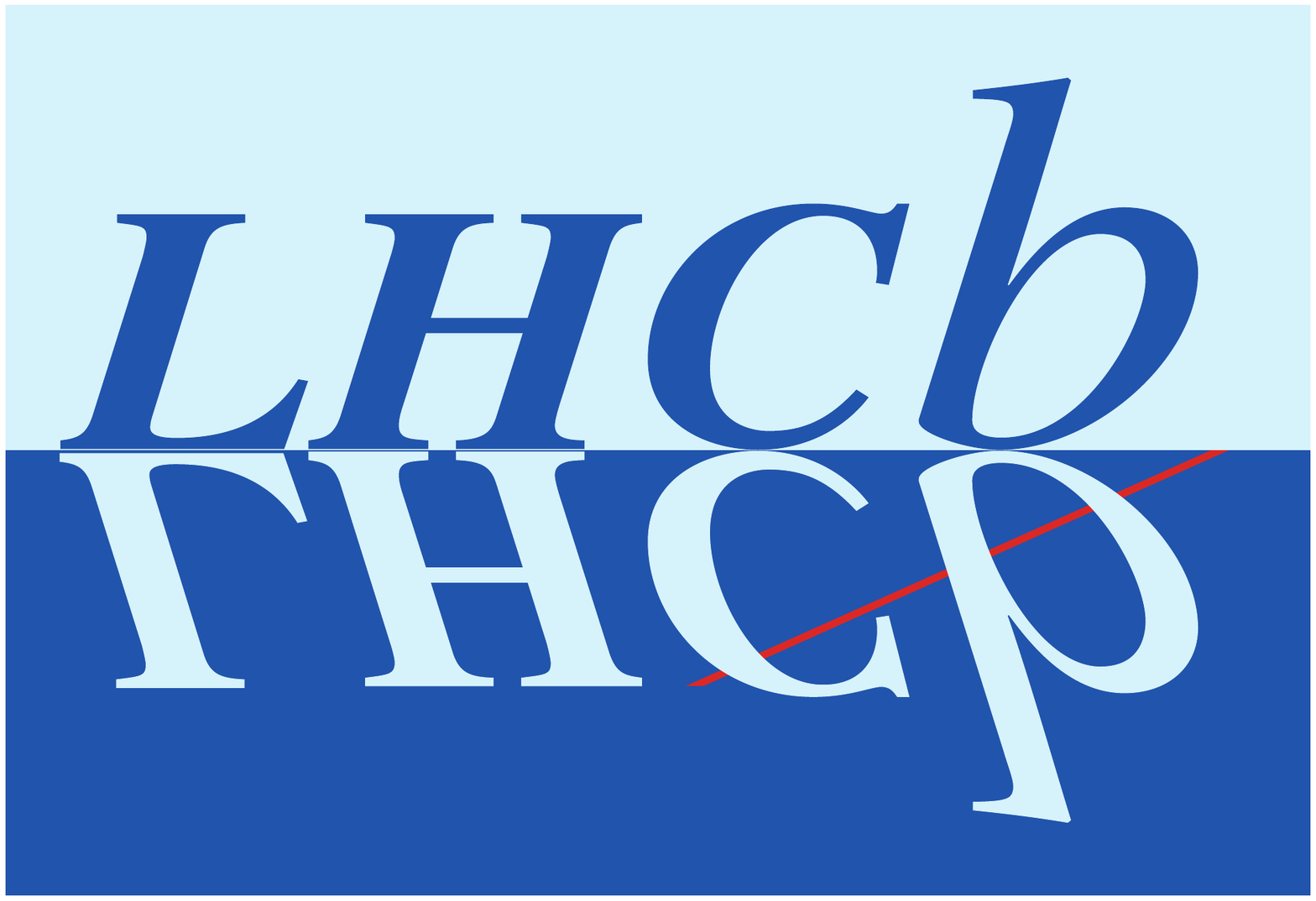}} & &}%
{\vspace*{-1.2cm}\mbox{\!\!\!\includegraphics[width=.12\textwidth]{lhcb-logo.eps}} & &}%
\\
 & & LHCb-PUB-2017-017 \\  
 & & \today \\ 
 & & \\
\end{tabular*}

\vspace*{4.0cm}

{\normalfont\bfseries\boldmath\huge
\begin{center}
  \thetitle
\end{center}
}

\vspace*{2.0cm}

\begin{center}
  \theauthors
\end{center}

\vspace{\fill}


\begin{abstract}
  We have examined the effects of embedded pitch adapters on signal formation in n-substrate silicon microstrip sensors with data from beam tests and simulation. According to simulation, the presence of the pitch adapter metal layer changes the electric field inside the sensor, resulting in slowed signal formation on the nearby strips and a pick-up effect on the pitch adapter.  This can result in an inefficiency to detect particles passing through the pitch adapter region. All these effects have been observed in the beam test data.
\end{abstract}

\vspace*{2.0cm}



\vspace{\fill}

{\footnotesize 
\centerline{\copyright~CERN on behalf of the \lhcb collaboration, licence \href{http://creativecommons.org/licenses/by/4.0/}{CC-BY-4.0}.}}
\vspace*{2mm}

\end{titlepage}


\newpage
\setcounter{page}{2}
\mbox{~}

\cleardoublepage

\renewcommand{\thefootnote}{\arabic{footnote}}
\setcounter{footnote}{0}
\pagestyle{plain} 
\setcounter{page}{1}
\pagenumbering{arabic}

\section{Introduction}
\label{sec:intro}

The pitch between strips of silicon microstrip sensors and the pitch between input channels of the application-specific integrated circuits (ASICs) used for readout can often differ by large amounts. One solution is to use external adapters such as metal traces on a glass substrate, but this presents its own difficulties for the construction of large detectors.  While the required wire bond geometry is consistent with industry standards, the number of bonds needed doubles. Moreover, additional space and material is introduced, which may be difficult to accommodate in densely packed silicon detectors.

Another solution to this problem is to use embedded pitch adapters built into the silicon sensors themselves.  During sensor fabrication a second metal layer of traces with bond pads designed to match the readout ASIC pitch is placed on top of the silicon dioxide passivation.  The AC-coupled metal strips of the first layer are directly connected with metal vias to this second layer above. 

This technology has been studied for the silicon tracker upgrade for the ATLAS experiment~\cite{Ullan2013178} and the Upstream Tracker (UT) upgrade for the LHCb experiment~\cite{LHCb-TDR-015}.
In the studies of the p-substrate sensor prototypes designed for the ATLAS experiment it was shown that the pitch adapters have an effect on the measured capacitance between strips~\cite{Ullan2016221}. However, no evidence of ``pick-up'' (creating spurious signals on far away strips) or of signal loss on nearby strips was found.  Sensors with increased strip segmentation utilizing routing lines in the first metal layer have been studied for upgrades to the CMS experiment~\cite{BOER2015154}.  In these sensors, charge pick-up on the routing lines and corresponding loss of signal on the nearby sensor strips was observed.

\begin{figure}[htb]
\begin{subfigure}{0.5\textwidth}
 \twgraphic{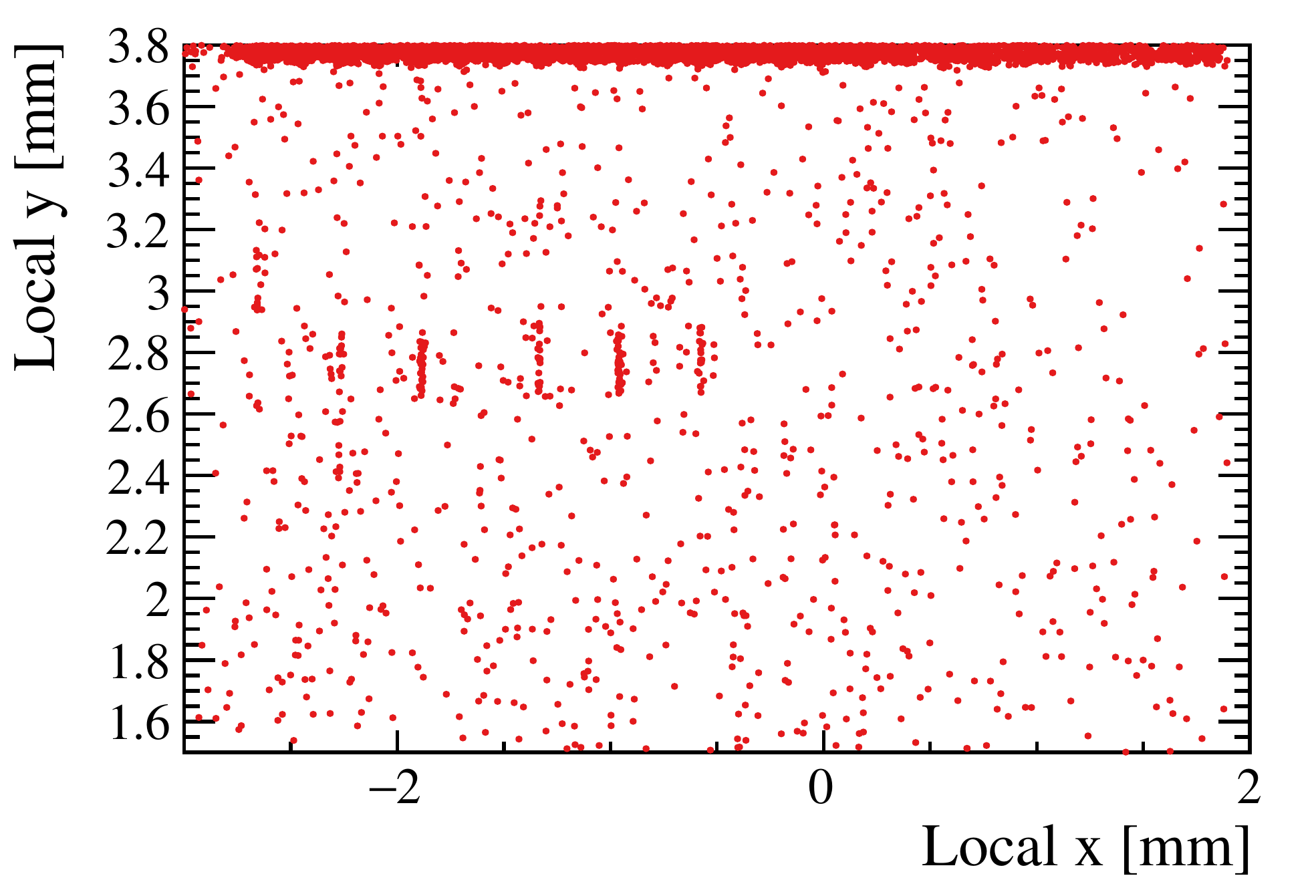}
  \caption{n-substrate sensor\label{fig:xray_ntype}}
\end{subfigure}
\begin{subfigure}{0.5\textwidth}
\twgraphic{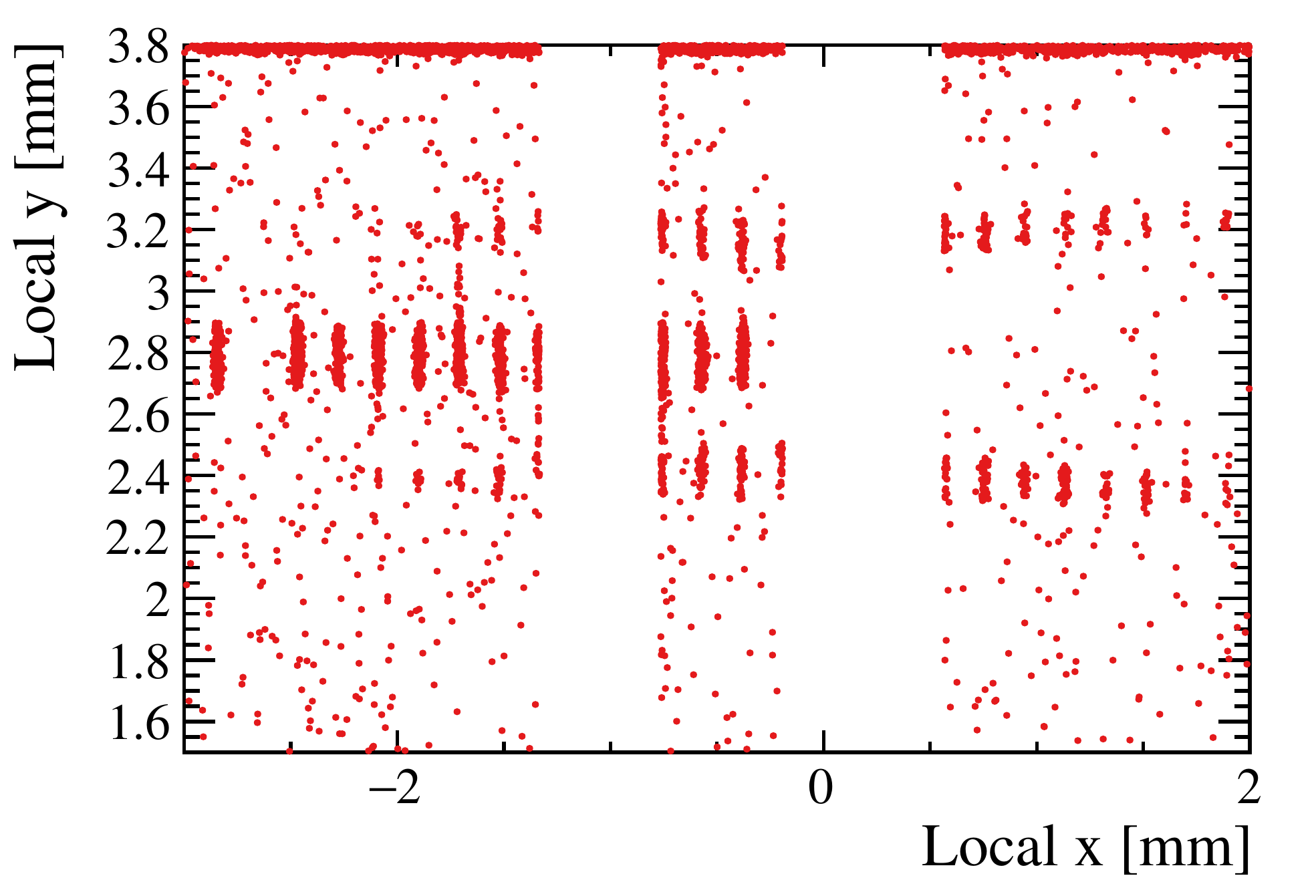}
\caption{p-substrate sensor\label{fig:xray_ptype}}
\end{subfigure}
\caption{ Positions on two sensors of charged particles for which a corresponding signal is not found in the sensor.  The inefficient band at the top corresponds to the edge of the sensor active area.  The pattern of inefficiency is confined to the pitch adapter region of the sensor and only in the region between two strips.  The n-type sensor (\protect\subref{fig:xray_ntype}) shows a small inefficiency near certain bond pads, while the p-type sensor (\protect\subref{fig:xray_ptype}) shows a much stronger effect.  The vertical bands with no missed hits are regions removed from analysis due to strips flagged as bad. \label{fig:xray}}
\end{figure}

During research and development for the UT, we observed a loss of efficiency localized in the pitch adapter (PA) region in both n- and p-substrate sensors.  Maps of the locations of passing charged particles that did not leave a reconstructed signal in the prototype sensor are shown in \cref{fig:xray}.  The inefficiency maps out the location of the embedded pitch adapter on these sensors.

In this paper, we focus on the effect in n-substrate sensors, using prototypes designed for the LHCb upgrade.  We have simulated the effects of the PA on signal formation in the sensor to understand the cause of the observed effects.  We then confirm further predictions of the simulation regarding the effect of design geometry and time evolution on the pick-up effect. 

\subsection{Sensor description}

The sensors used are miniature ($\SI{1.8}{\centi\meter}\times\SI{1.4}{\centi\meter}$) silicon microstrip sensors produced by Hamamatsu Photonics~\cite{hamamatsu}.  The silicon is \SI{320}{\micro\meter} thick and of p$^+$-in-n type.  Each sensor has 64 strips with a pitch of \SI{190}{\micro\meter}.  The sensor's AC-coupled strips are connected to the readout ASICs with a pitch of approximately \SI{80}{\micro\meter} through the use of an embedded pitch adapter.

\begin{figure}[htb]
  \includegraphics[width=\textwidth,angle=180]{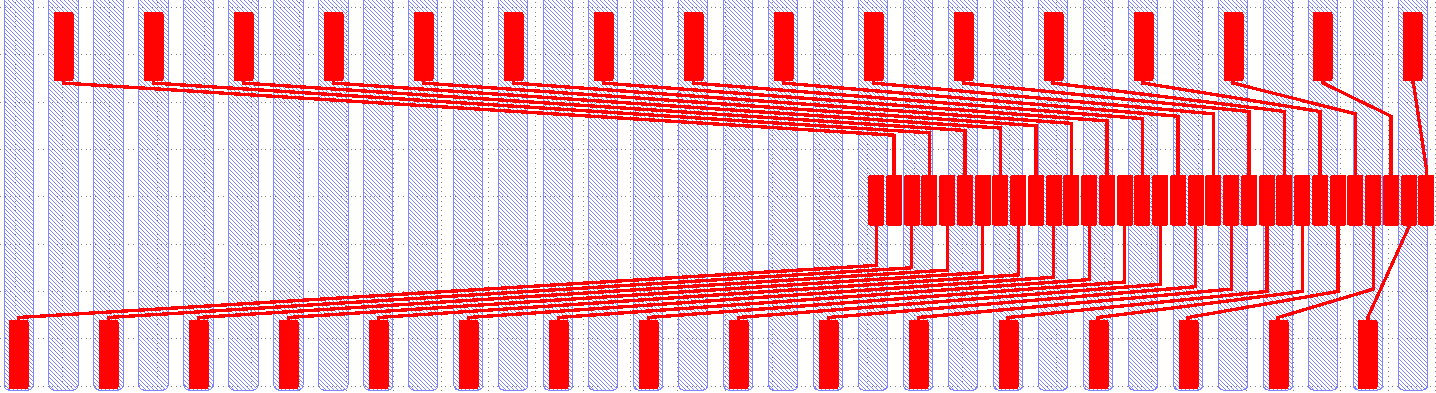}
  \caption{Zoomed in view of the sensor metalization layers, showing one-half of the pitch adapter.  The hatched purple vertical bands show the metal strips above each silicon strip.  Thirty-two channels are connected via the second-metal layer shown in solid red, to the closely spaced bond pads in the center.  The other thirty-two channels are similarly connected with a mirror-image design not pictured. \label{fig:sensor}}
\end{figure}

The design of the pitch adapter is pictured in \cref{fig:sensor}. Metal traces with a width of approximately \SI{10}{\micro\meter} connect each channel to a bonding pad \SI{65}{\micro\meter} wide (perpendicular to the strips) and \SI{210}{\micro\meter} long (parallel to the strips).  The distance between the metal traces in the densest region varies from \SIrange{10}{20}{\micro\meter}.   These traces and bonding pads are located on top of the active sensor area.  The metal layer of the pitch adapter is separated from the first metal layer containing the strip AC-metal by a layer of silicon dioxide that is \SI{2}{\micro\meter} thick.  The sensor's metal strips are \SI{123}{\micro\meter} wide, with a gap between of \SI{67}{\micro\meter}.

The analysis in this paper focuses on the bond pad region, where the larger metal area is expected to cause stronger effects.  The bond pads almost completely bridge the inter-strip region when they are centered relative to the strips below.  In other cases, the bond pad is shifted to one side or the other, partially or fully overlapping the metal strip.

\section{Simulation}
\label{sec:simulation}

\begin{figure}[htb]
\begin{center}
  \includegraphics[width=0.5\textwidth]{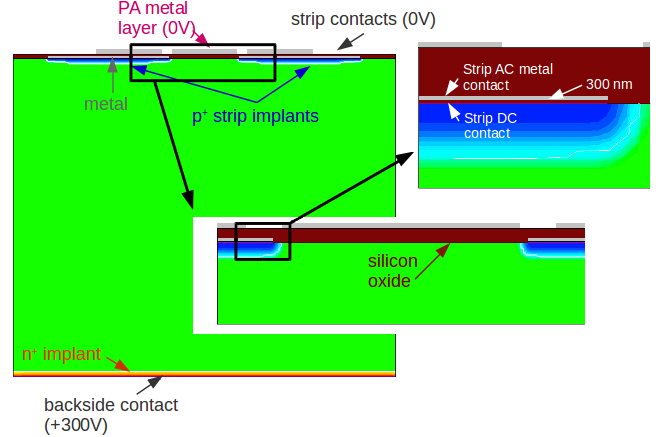}
\end{center}
  \caption {Two dimensional model of the sensor used for the simulation.  The successive insets zoom in on the details of the
    implant and metalization geometry near the sensor
    surface. The metal layers of the strip AC-metal and pitch adapter have had their thickness exaggerated to improve the visualization.\label{fig:scheme_Interstrip}}
\end{figure}

We simulate the operation of the sensor in the pitch adapter region using the TCAD Sentaurus package, part of Synopsys software~\cite{Synopsys}. We use a two dimensional simulation of the full sensor in a slice perpendicular to the strips, including two implant strips and a metal strip to model the pitch adapter bond pad.  The geometry of the model is shown in \cref{fig:scheme_Interstrip}.

\subsection{Model description}

The exact parameters of the sensors manufactured by Hamamatsu are not known, therefore the choice of many parameters for the simulation was inspired by studies of similar silicon sensors produced by Hamamatsu for CMS~\cite{EberThesis}. We assume the bulk to be doped with phosphorus (n-type) with a concentration of \SI{3e12}{\per\cubic\centi\meter}.  The strips implants are doped with boron (p$^+$) with a much higher concentration of \SI{1e19}{\per\cubic\centi\meter}.  A Gaussian error function is used for the implant doping profile with a depth of \SI{1}{\micro\meter} and a width of \SI{1}{\micro\meter}.  Variations of the implant profile 
depth and width in ranges 1-\SI{5}{\micro\meter} and 0.5-\SI{1.5}{\micro\meter}, respectively,
have shown no significant effects on the results of the simulation.
The bottom side of the bulk is doped with phosphorus (n$^+$) with a concentration of \SI{0.5e19}{\per\cubic\centi\meter} to achieve ohmic contact with the metal backplane. A density of positive charges trapped at the boundary of the silicon and silicon dioxide of \SI{1e11}{\per\square\centi\meter} is assumed for unirradiated sensors. To emulate the effect of irradiation, the charge density is increased to its saturation value of \SI{3e12}{\per\square\centi\meter}.  These numbers are based on typical values for silicon--silicon dioxide boundaries~\cite{mos,JSCHWANDT2017159}, but have not been measured on the tested sensors.  It is possible that the tested sensors start with a lower charge density than \SI{1e11}{\per\square\centi\meter} when unirradiated, but changes to the qualitative behavior of the sensors are not observed below this value.

We chose for the width of the simulation region to be twice the pitch of \SI{190}{\micro\meter}, and thus it includes two p$^+$ strips each \SI{120}{\micro\meter} wide. The p$^+$ implants are connected to electrical ground via \SI{1.5}{\mega\ohm} resistors and AC coupled to the metal strips placed at \SI{300}{\nano\meter} above them in the silicon dioxide layer. When simulating the pitch adapter pads, a second metal layer on top of the silicon dioxide is held at electrical ground to represent the  \SI{65}{\micro\meter} wide pitch adapter pads.  Three pads are included with a separation of \SI{10}{\micro\meter}; the center pad is  placed over the inter-strip region; its edge is at a distance of \SI{2.5}{\micro\meter} from the edge of the strip. A positive high voltage bias of \SI{300}{\volt} (the depletion voltage is approximately \SI{250}{\volt}) is applied to the backside contact. An adaptive meshing with cell dimensions from $\SI{20}{\micro\meter}\times\SI{50}{\micro\meter}$ to $\SI{1}{\micro\meter}\times\SI{1}{\micro\meter}$ is used.

It was shown in Ref.~\cite{EberThesis} that the TCAD Sentaurus default
values for the most important parameters (permittivity, electron
affinity, band gap energy, permittivity of silicon dioxide) are well
justified and therefore were used in this study. Though the default
generation and recombination times of electron and holes are lower
than those of very clean silicon, we do not expect it to have a
significant influence on the results of this investigation.  The
simulation of charge carrier mobility accounts for velocity saturation
at high fields and doping dependency.  The generation and
recombination of charge carriers is implemented according to the
Shockley-Read-Hall recombination model.  At the boundary an absence of
electric field and carrier current perpendicular to the boundary are
required~\cite{BoundaryPennicard}. 


\subsection{Results of the simulation}

\begin{figure}[tb]
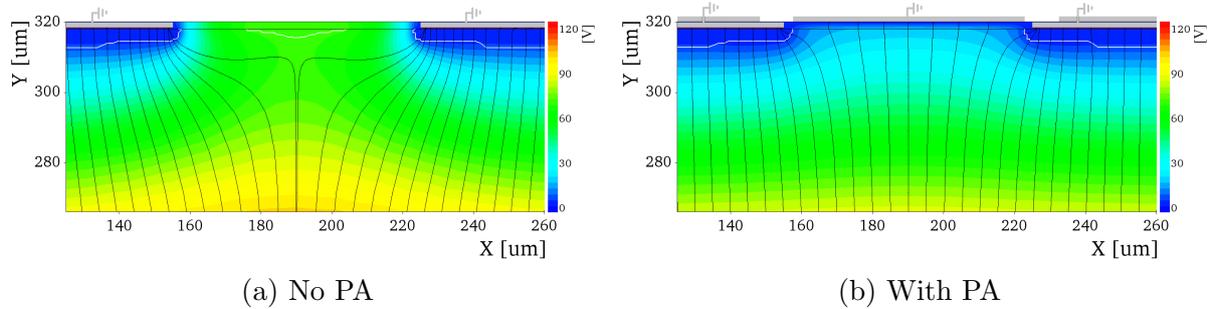

  \begin{subfigure}{0.5\textwidth}
    \twgraphic{simulation/INTmap_potential_noPA}
    \caption{No PA}
  \end{subfigure}
  \begin{subfigure}{0.5\textwidth}
    \twgraphic{simulation/INTmap_potential_withPA}
    \caption{With PA}
  \end{subfigure}
  \caption {Simulated map of the electrostatic potential in the inter-strip
    region at \SI{300}{\volt} bias. The left figure corresponds to a
    configuration without pitch adapter pads, while the right one
    includes them. The white line represents the boundary of the
    depletion region. The metal layers of the strip AC-metal and pitch adapter have had their thickness exaggerated to improve the visualization.\label{fig:INTmap}}
\end{figure}

\begin{figure}[tb]
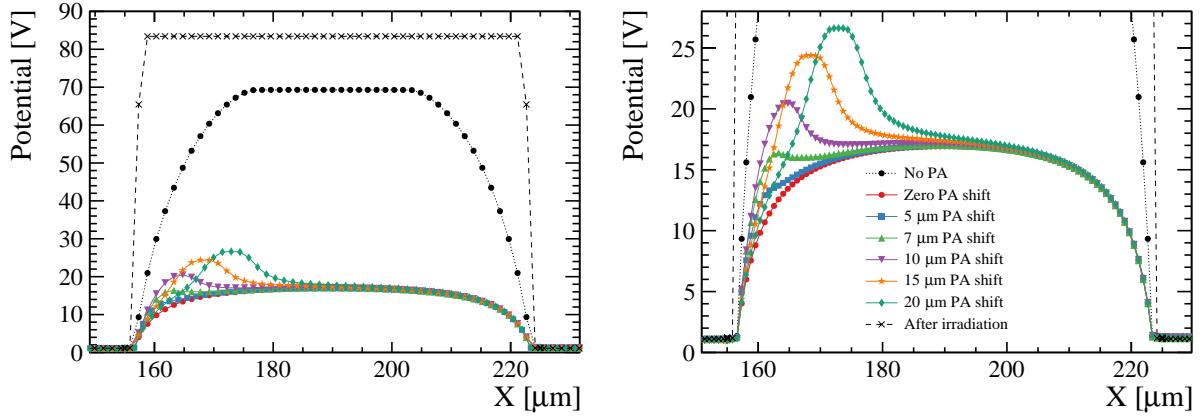


  \begin{subfigure}{0.5\textwidth}
    \twgraphic{simulation/Potential_PAshift1}
  \end{subfigure}
  \begin{subfigure}{0.5\textwidth}
    \twgraphic{simulation/Potential_PAshift2}
  \end{subfigure}
 \caption {Simulated electrostatic potential profiles at the silicon--oxide
   boundary in the inter-strip region for various PA pad positions and
   without PA.  Zero PA shift centers the PA metal pad between two strips, \SI{2.5}{\micro\meter} from either strip; each shift increases the distance on one side. Right figure shows zoomed profiles. The dotted line
   with black circles represents the case without any PA pads. The
   dashed line with crosses represents the case of an irradiated sensor
   modeled by using saturation value of
   \SI{3e12}{\per\square\centi\meter} for oxide charge density
   (irrespective of presence and position of PA pads).
 \label{fig:PotentialProfile}}
\end{figure}

The electrostatic potential maps from simulation are shown in \cref{fig:INTmap} for configurations with and without the pitch adapter metal layer. The presence of the pitch adapter layer changes the electric field in the inter-strip region close to the silicon surface.  The charge carriers (holes) are attracted to the silicon-oxide surface and only then travel towards the strip implants. A shift in the PA pad position with respect to the middle of the inter-strip region results in a remarkable effect. An electrostatic potential at the silicon-oxide boundary rises near the strip opposite to the shift direction, producing a barrier as shown in \cref{fig:PotentialProfile}.  The gradient of the potential in the $x$ direction increases the coupling to the second metal and decreases the charge collection.

\begin{figure}[tb]
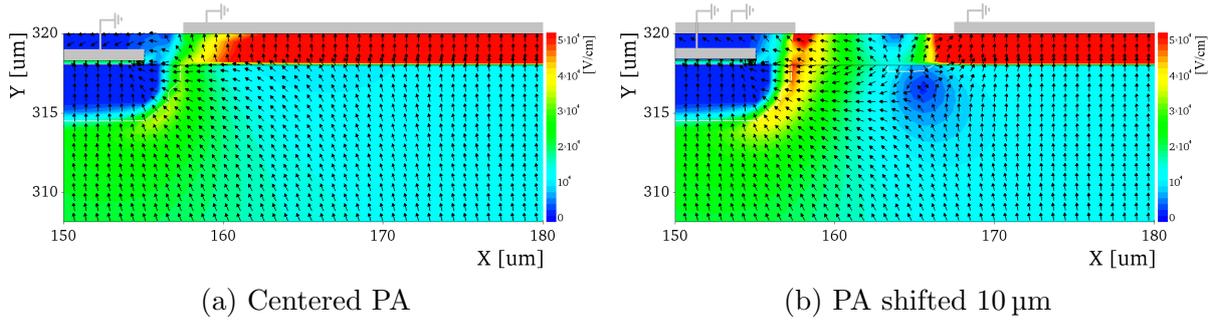

  \begin{subfigure}{0.5\textwidth}
    \twgraphic{simulation/Ox2_PA2_Shift0_EField_zoom}
    \caption{Centered PA\label{fig:efield_cent}}
  \end{subfigure}
  \begin{subfigure}{0.5\textwidth}
    \twgraphic{simulation/Ox2_PA2_Shift10_EField_zoom}
    \caption{PA shifted \SI{10}{\micro\meter}\label{fig:efield_shift}}
  \end{subfigure}
  \caption{ Simulated electric field maps in the region near the silicon surface close to a strip implant.  The color scale represents the field strength and the printed arrows the direction.  In (\protect\subref{fig:efield_cent}), the PA is centered.  In (\protect\subref{fig:efield_shift}), the PA is shifted \SI{10}{\micro\meter}, creating a zero field region near the surface that blocks charge moving towards the strip on the left side.  The metal layers of the strip AC-metal and pitch adapter have had their thickness exaggerated to improve the visualization.\label{fig:efield}}
\end{figure}

This potential creates a region of zero electric field as shown in \cref{fig:efield}.  The field prevents the charge carriers from reaching the nearby strip for a
timescale long compared to the usual readout.  This results in a signal loss on the strip and signal pick-up on the corresponding PA pad, where the induced charge is longer lived.
In case of saturated charge density in the silicon-oxide interface (corresponding to an irradiated sensor) the effect of the PA pads is suppressed and the resulting potential is the same as without PA pads.

\begin{figure}[tb]
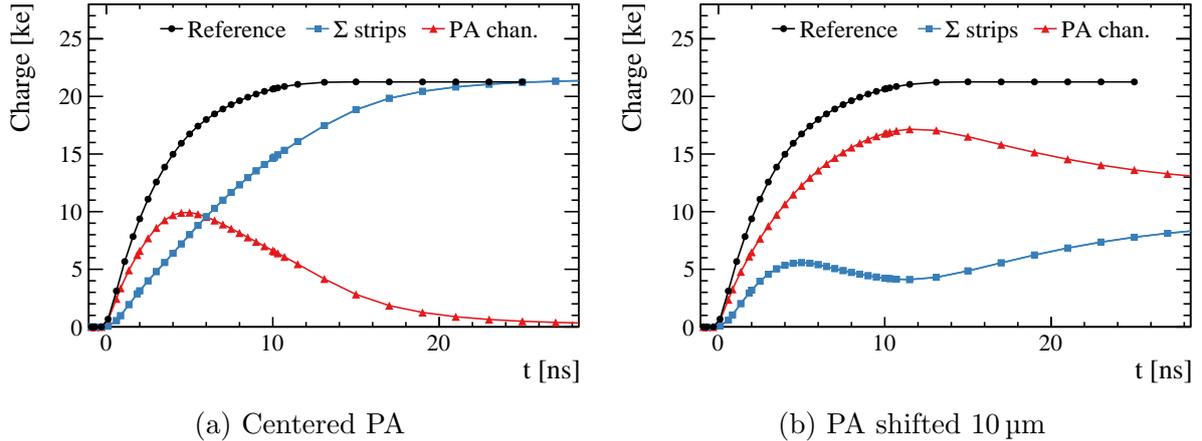

  \begin{subfigure}{0.5\textwidth}
    \twgraphic{simulation/MIP_noShift}
    \caption{Centered PA \label{fig:MIP_noshift}}
  \end{subfigure}
  \begin{subfigure}{0.5\textwidth}
    \twgraphic{simulation/MIP_Shift10}
    \caption{PA shifted \SI{10}{\micro\meter} \label{fig:MIP_shift}}
  \end{subfigure}
  \caption {Simulated charge appearing on the strip AC and PA contacts versus time since MIP hit in the inter-strip region \SI{5}{\micro\meter} closer to one of the strips. The reference curve corresponds to charge on the strip AC contact in response to a MIP passing under the strip.  To account for charge sharing, the sum of the charge collected on both strips is used.}
 \label{fig:MIP}
\end{figure}

To investigate signal pulse shapes, we simulate hits of minimum ionizing particles (MIP) penetrating the sensor perpendicularly to its plane, near to the halfway point between two strips. The MIP ionization density is tuned to produce approximately \SI{2e4} electron-hole pairs in the bulk.  The total charge induced on the strips' coupling capacitors and on the pitch adapter metal (before any shaping) as a function of the integration time are shown in \cref{fig:MIP} for two variants of PA position. The reference shape corresponds to the pulse from a particle passing directly under a strip. When the PA pads are centered with respect to the inter-strip region, slower collection on the nearby strips is observed with respect to the reference while a pick-up signal on the PA emerges with a peaking time of approximately \SI{5}{\nano\second} (\cref{fig:MIP_noshift}). In the case where the PA pads are shifted \SI{10}{\micro\meter}, resulting in a distance to the farther strip of \SI{12.5}{\micro\meter}, the pick-up signal on the PA becomes more similar to the reference while the signal on the strips is reduced significantly.  All the charge eventually reaches the strips on the order of microseconds.  This case is likely to result in a localized region of inefficiency.

\begin{figure}[tb]
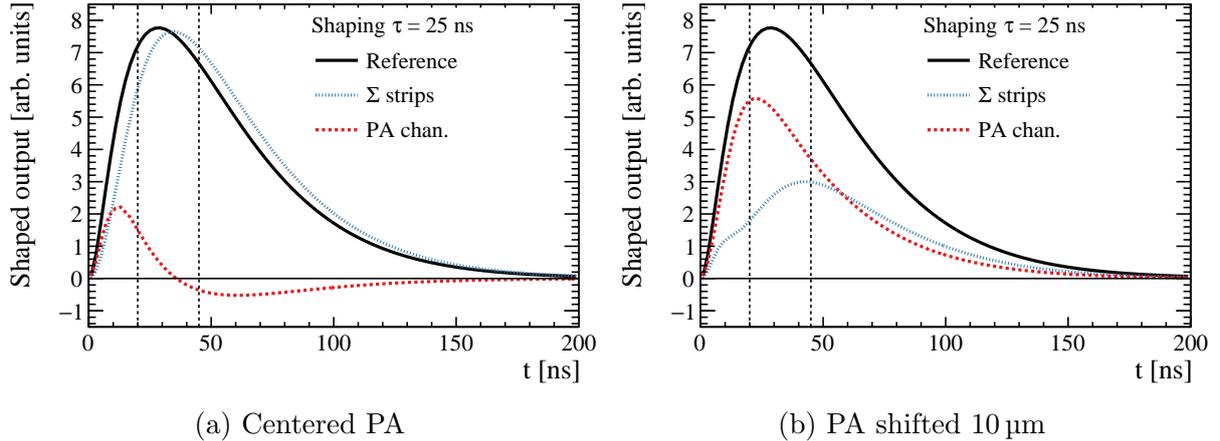

  \begin{subfigure}{0.5\textwidth}
    \twgraphic{simulation/pulse_noShift}
    \caption{Centered PA \label{fig:pulse_noshift}}
  \end{subfigure}
  \begin{subfigure}{0.5\textwidth}
    \twgraphic{simulation/pulse_Shift10}
    \caption{PA shifted \SI{10}{\micro\meter} \label{fig:pulse_shift}}
  \end{subfigure}
  \caption{Simple pulse shapes produced from the simulated charge appearing on the strip AC and PA contacts versus time since MIP hit in the inter-strip region \SI{5}{\micro\meter} closer to one of the strips. The reference curve corresponds to the one in \cref{fig:MIP}.  To account for charge sharing, the sum of the charge collected on both strips is used.  The vertical lines denote a \SI{25}{\nano\second} interval, motivated by the time window imposed by the data acquisition system on data from the beam test.}
 \label{fig:pulses}
\end{figure}

In order to qualitatively compare with data from beam tests, a simple shaping function of the form:
\begin{equation}
  W( t ) = \frac{t}{\tau}e^{-t/\tau},
\end{equation}
has been applied to the charges, where $\tau$ is a shaping time of \SI{25}{\nano\second}.  The results are shown in \cref{fig:pulses}.  When the PA is centered, the slowed development of the signal on the nearby strips should be visible as a shift in the time of the pulse peak.  The pick-up signal on the PA resembles a very large cross-talk signal.  When the PA shifts, the signal on the nearby strips is greatly reduced.  The PA then sees a pulse similar in strength to a real signal.

\section{Beam test}
\label{sec:testbeam}

The sensors have been tested using beams delivered from the CERN SPS to the H8 beam line in the CERN North Area.  This beam delivers positively charged hadrons with a momentum of \SI{180}{\giga\evolt}.
  The width of the beam is approximately \SI{1}{\centi\meter}, allowing the beam to simultaneously illuminate the pitch adapter area and the edges of the sensor active area.

\subsection{Experimental setup}

The sensor, referred to as the device under test (DUT), is placed at the center of the Timepix3 telescope; an earlier iteration of this telescope is described in Ref.~\cite{Akiba:2013yxa}.  This telescope consists of 8 Timepix3 pixel modules used to reconstruct the trajectory of passing charged particles.  These tracks are extrapolated to where the DUT is placed with a pointing resolution of approximately \SI{2}{\micro\meter}.  This allows for a precise determination of the impact points of the beam particles with respect to the sensor structures like the strip implants and pitch adapter.  The beam impacts the DUT perpendicularly.

The sensors are read out using the Beetle ASIC~\cite{Lochner:2006vba}.  This chip outputs the signals on every channel as an analog pulse with a shaping time on the order of tens of \si{\nano\second}.  These signals are processed using the MAMBA data-acquisition (DAQ) board produced by Nuclear Instruments~\cite{mamba}.
This DAQ system samples and digitizes the analog pulse at a fixed time that can be moved in \SI{25}{\nano\second} increments.  The beam particles arrive asynchronously with the \SI{40}{\mega\hertz} clock used for readout.  This allows for the study of particles arriving at all times within the window.

A copy of the scintillator signal used for triggering the DUT is fed into the data stream of the Timepix3 telescope.  A shared time-stamp is used to match tracks reconstructed from the Timepix3 hits with the DUT data for the corresponding trigger.

To test the effects of radiation, sensors were irradiated at the IRRAD~\cite{irrad} facility using protons with a momentum of \SI{24}{\giga\eV}.  Dosimetry measurements were used to determine the particle fluence in the \SI{3}{\milli\meter} band surrounding the pitch adapter.  Sensors with six different fluences ranging from zero to \SI{2e13}{\neq} were used in our tests.  The latter fluence represents the maximum expected for UT sensors with these pitch adapters after 50\invfb of data collected in LHCb.

Miniature sensors from this study were found to have signal-to-noise ratios of at least 20.  Further information on similar beam tests conducted with prototype sensors from the same project may be found in Ref.~\cite{Abba:2137551}.

\subsection{Data sample and analysis}

In order to isolate the effects of the pitch adapter and compare to simulation, specific beam events are chosen in which the passing particle goes through the bond pads of the pitch adapter.  Only events in which a single track is found in the Timepix3 telescope are selected.  Using the coordinates from tracks extrapolated to the DUT position, the side and top edges of the sensor are determined.  The position of the track in the local $x$ (perpendicular to each sensor strip) and local $y$ (parallel to each strip) directions is then calculated.
Results from multiple runs with the same sensor, taken in a short period of time, are combined together, but the position calibration is repeated for each run independently.

Fourteen strip pairs have been identified from the sensor design for which a single PA bond pad is found over the inter-strip region.  Some of these pads are centered between the two strips, and some are shifted towards one side or the other.  Tracks passing within \SI{19}{\micro\meter} of the center of the inter-strip region, at the correct range of local $y$ values to be under the bond pad are identified.  These 14 rectangular regions together make up the ``PA region''.  For each strip pair, the channel that is connected to the PA bond pad in that region is also identified.  The output for these three channels are then studied together to analyze the effect of the pitch adapter.  

For comparison, a control region is defined using the inter-strip regions of the same 14 strip pairs.  The control region requires the track to pass away from the PA region by requiring values of local $y$ farther down the sensor.  The signal response in the control region has been validated to be free of pitch adapter effects by comparing it with tracks that pass directly underneath a single strip.

For each event in the sample, the ADC counts for the two channels corresponding to the nearby strips as well as the channel connected to the PA bond pad are studied.  The ADC values used are taken after pedestal and common-mode noise subtraction are performed.  Events in which one of the three strips (nearby or PA connected) is flagged as noisy or dead are removed from the analysis.

The data output also includes a timing value for the event.  The \SI{25}{\nano\second} readout window is divided into ten \SI{2.5}{\nano\second} time bins, corresponding to different trigger arrival times relative to the \SI{40}{\mega\hertz} readout clock.  The data in each time bin is therefore sampled from the analog Beetle output pulse at different times relative to the start time of the pulse.  This allows us to reconstruct the shape of the output pulse inside the \SI{25}{\nano\second} window.

\section{Results from beam test}
\label{sec:nresults}

\begin{figure}[htb]
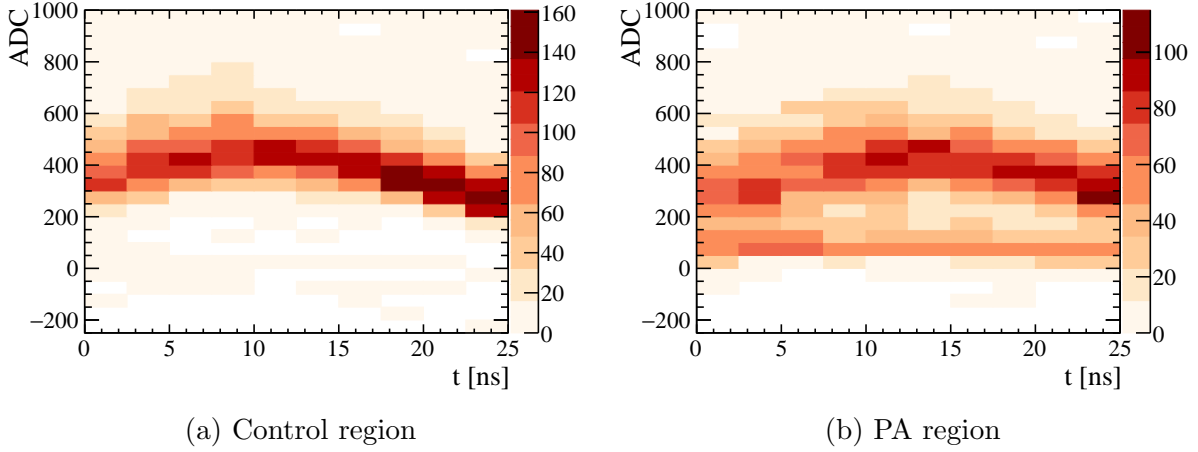

  \begin{subfigure}{0.5\textwidth}
    \twgraphic{ntyperesults/ctl_sum_btwn}
    \caption{Control region\label{fig:npulse2d_ctl}}
  \end{subfigure}
  \begin{subfigure}{0.5\textwidth}
    \twgraphic{ntyperesults/pa_sum_btwn}
    \caption{PA region\label{fig:npulse2d_pa}}
  \end{subfigure}
  \caption{Sum of the ADC values for the two strips near the track location versus time for an unirradiated sensor.  The tracks in the control region are shown in (\protect\subref{fig:npulse2d_ctl}) and in the PA region in (\protect\subref{fig:npulse2d_pa}). \label{fig:npulse2d}}
\end{figure}

Two dimensional views of the sampled ADC versus time are shown in \cref{fig:npulse2d} for the PA and control regions.  The control region shows a single clear peak in the signal pulse as a function of time.  Two qualitatively different signal patterns are clear in the data from the PA region.  In some events, there is a clear signal peak on the two closest strips similar to the control region, while in others the nearby strips show a value close to zero.  In the latter case, the resulting data would often fail clustering requirements to separate signal from noise and the hit would be missed.  These two classes of events are separated by requiring the sum ADC on the two nearby strips to be less than or greater than 150.  Since the signal is split almost evenly between the two strips, this is approximately equivalent to a cut of three times the noise applied to each strip separately.

To reconstruct the pulse shape for each class of events, the peak ADC value is first determined separately for each time bin.  The pulse is reconstructed separately for nearby strips and the PA connected strip.  For the nearby strips, the ADC values of the two channels are added together before finding the peak.  Three different methods for peak finding are used:
\begin{itemize}
\item For good signals ($\text{ADC} > 150$) on the nearby strips, the ADC distribution is fitted with a Landau convolved with a Gaussian.  The peak position is taken from the most probable value.
\item False ``signals'' on the PA connected strip show a different ADC distribution without the characteristic high-side tail of a Landau distribution, but this may be due to the small number of recorded events. For these outputs, the mean value from a Gaussian fit is used to find the peak position.
\item For channels without clear signals (cross-talk and inefficient hits), the mean of the ADC values is taken as the peak position.
\end{itemize}

\begin{figure}[htb]
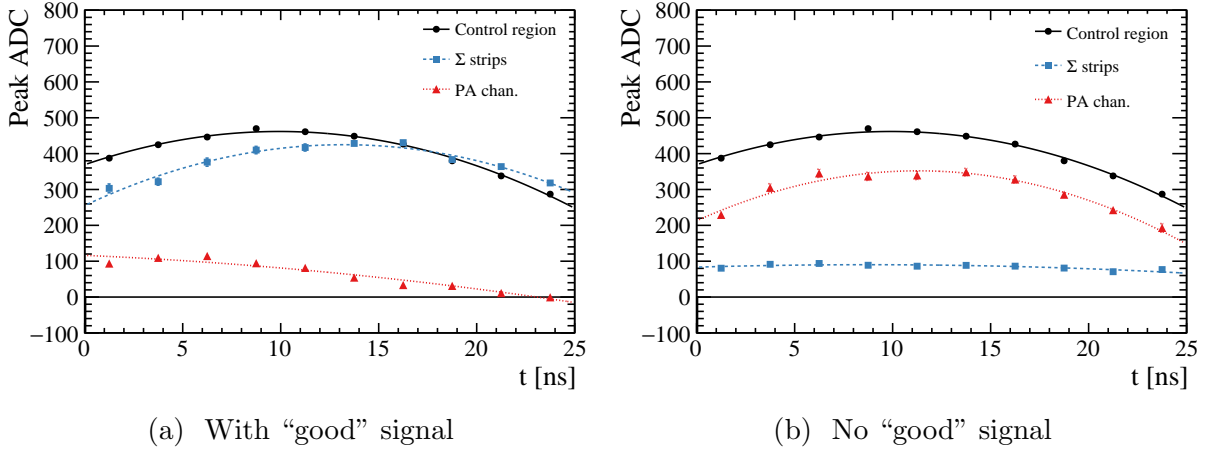

  \begin{subfigure}{0.5\textwidth}
    \twgraphic{ntyperesults/pa_pulse}
    \caption{\label{fig:npulse_good} With ``good'' signal }
  \end{subfigure}
  \begin{subfigure}{0.5\textwidth}
    \twgraphic{ntyperesults/far_pulse}
    \caption{\label{fig:npulse_bad} No ``good'' signal }
  \end{subfigure}
  \caption{Reconstructed pulse shapes within the \SI{25}{\nano\second} readout window for particles passing halfway between two strips.  The pulses in the control and PA regions represent the sum of the output on the two nearby channels.  In (\protect\subref{fig:npulse_good}), a good signal is found on the nearby strips when the particle lands in the PA region; in (\protect\subref{fig:npulse_bad}) a good signal is not found.
    \label{fig:npulse}}
\end{figure}

Once the peak value in each bin has been determined, the resulting points are fit with a quadratic function to qualitatively describe the shape of the pulse within the \SI{25}{\nano\second} window.  The results are shown in \cref{fig:npulse}.  The two cases of a ``good'' signal  ($\text{ADC} > 150$) found or not found on the nearby strips are compared.  In the pitch adapter region, good signals are slowed in time and reduced in magnitude relative to the control region pulse; the peak occurs approximately \SI{5}{\nano\second} later.  Simultaneously, a large cross-talk like signal appears on the PA connected strip, already decreasing in the sampled time window.  These results match qualitatively the expectation from simulation.

\begin{figure}[htb]
  \begin{center}
  \includegraphics[width=0.5\textwidth]{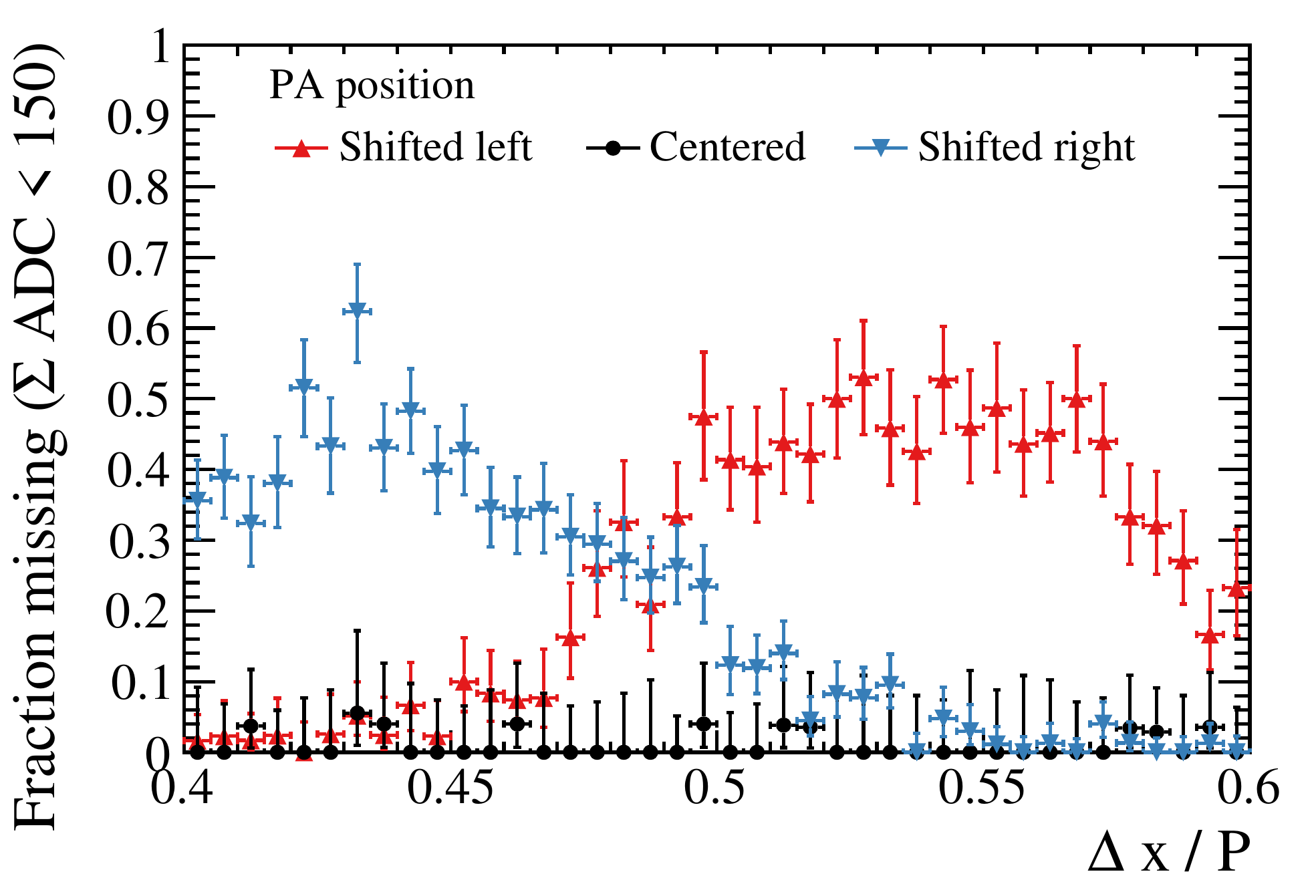}
  \end{center}
  \caption{Fraction of particles with missing signals on the nearby channels as a function of the inter-strip position for particles passing through the PA region.  $\Delta x/P = 0.5$ corresponds to halfway between the two strips.  Three groups of channels are shown, corresponding to cases in which the PA bond pad is either centered between the two strips, or shifted towards one side or the other.
    \label{fig:ntype_dx_miss}}

\end{figure}

When the signal on the nearby strips is lost, a pulse similar in shape to a good signal is instead found on the PA connected strip.  The nearby strips show only a small signal above zero.  The source of these events can be determined by further subdividing the data sample.  For two of the 14 strip pairs, the PA bond pad is centered above the inter-strip region.  For six pairs it is shifted by approximately \SI{10}{\micro\meter} towards one side, and the other six towards the other side.  In \cref{fig:ntype_dx_miss}, the fraction of missing hits for each of these three groupings is shown as a function of the inter-strip position normalized by the strip pitch ($\Delta x /P$).  For the centered bond pads, no inefficiency is found.  The shifted bond pads show an inefficiency in the opposite side of the inter-strip region which is consistent with the expectation from simulation.  The local inefficiency reaches approximately 50\% at the highest.  The overall effect averaged over the sensor is small since the region of inefficiency is limited to this small area.

\begin{figure}[htb]
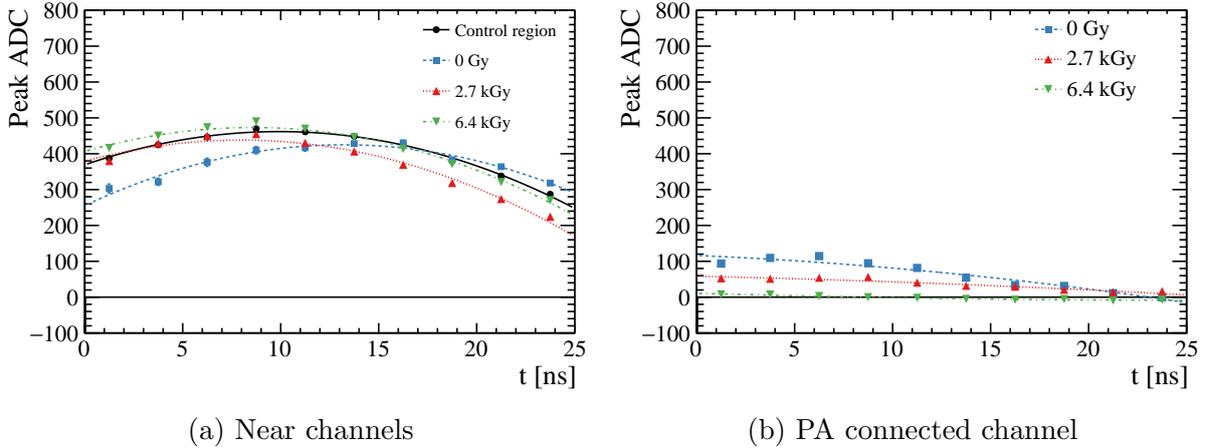

  \begin{subfigure}{0.5\textwidth}
    \twgraphic{ntyperesults/pa_pulse_irrad}
    \caption{Near channels\label{fig:npulse_near_irrad}}
  \end{subfigure}
  \begin{subfigure}{0.5\textwidth}
    \twgraphic{ntyperesults/pa_far_irrad}
    \caption{PA connected channel\label{fig:npulse_far_irrad}}
  \end{subfigure}
  \caption{Pulse shapes for three sensors with different radiation doses in the silicon dioxide layer.  The radiation doses are calculated based on ionization in the silicon dioxide by the protons used for irradiation.  As irradiation increases, the PA coupling effect is reduced, both in the delay in the signal observed in the nearest two channels (\protect\subref{fig:npulse_near_irrad}) and in the PA connected channel signal (\protect\subref{fig:npulse_far_irrad}). \label{fig:npulse_irrad}}
\end{figure}

We find that the effects of the PA are reduced with irradiation, as shown in \cref{fig:npulse_irrad}.  By a dose of \SI{6.4}{\kilo\gray} in the silicon dioxide layer, no PA coupling effects are found.  At lower levels of irradiation, the effects on the nearby strip signals and on the PA connected channel are reduced but not totally eliminated.  With the measurements available on these sensors, we are unable to determine precisely the level of charge build-up in the oxide layer as a function of dose.

\section{Conclusions}
\label{sec:concl}

We have observed effects on signal collection due to the presence of embedded pitch adapters over the active sensor area in n-substrate silicon microstrip sensors.  Some pitch adapter geometries result in a loss of signal on the nearest strips when a charged particle passes underneath the second metal layer.  This results in an inefficiency for particle detection after thresholds are applied to reject noise.  Simultaneously, a signal-like pulse is found on the channel connected to the pitch adapter element.

Using simulation, we are able to understand these effects in more detail.  The presence of the second metal layer over top of the region between two implant strips alters the electric fields inside the bulk of the silicon.  When the metal layer is centered over the inter-strip region, this results in a slower time evolution of the signal pulse. We observe this change in timing in our data.  The channel connected to the pitch adapter experiences a large cross-talk signal, which we also observe.

When the pad in the second metal layer is shifted relative to the inter-strip region, a potential barrier can form that impedes the collection of signal over a time scale long enough to cause an inefficiency.  When this occurs, the pitch adapter connected channel outputs a larger pulse that is easily mistaken for a true hit.  The inefficient hits in our data are limited to cases in which the pitch adapter bond pad is shifted by approximately \SI{10}{\micro\meter}.

We have observed that these problems disappear with irradiated sensors. This could be explained by the build up of positive charge at the silicon dioxide--silicon interface, which is known to occur with irradiation~\cite{mos}. Simulation shows that a surface charge density around \SI{1e12}{\elementarycharge\per\centi\meter\squared} should suffice. It is plausible that the irradiation of the sensors used for the beam test has produced this charge density, although we were not able to directly measure this charge.

We have investigated in simulation  scenarios that may reduce the effects of the PA. We predict that doubling the thickness of the silicon dioxide layer from \SI{2}{\micro\meter} to \SI{4}{\micro\meter} should remove the inefficiency and pick-up signals we have observed.

All of these effects are localized to the small regions directly underneath pitch adapter metal and halfway between two strips.  For these sensors, we only observe problems near the larger bond pads; no significant effect is seen under the thin traces.  For all particles passing through other parts of the detector, we observe no effects on signal formation due to the pitch adapters.

Similar effects have also been observed in p-substrate sensors.  Detailed understanding of the effects in these sensors requires future beam test and simulation efforts.

\section{Acknowledgements}
\label{sec:ack}

We would like to express our gratitude to Hamamatsu for producing the sensors used in this study.  We thank the CERN accelerator departments for the excellent performance of the SPS which provided the beam used in our tests.  We also thank the VELO Timepix telescope group for the excellent instrument that made the precision studies possible.  The IRRAD facility at CERN also provided valuable help irradiating the tested sensors.    We acknowledge the support of the National Science Foundation for this work.

\setboolean{inbibliography}{true}
\bibliographystyle{LHCb}
\bibliography{main,LHCb-PAPER,LHCb-CONF,LHCb-DP,LHCb-TDR}

\end{document}